\documentclass{elsart}

\usepackage{natbib}

\usepackage{graphics}
\usepackage{graphicx}

\begin{document}

\begin{frontmatter}



  \title{ Virial statistical description of non-extensive hierarchical
    systems}

  \author{Daniel Pfenniger}

  \address{Geneva Observatory, University of Geneva, CH-1290 Sauverny.
    Switzerland} \ead{daniel.pfenniger@obs.unige.ch}

  \begin{abstract}
    In a first part the scope of classical thermodynamics and
    statistical mechanics is discussed in the broader context of
    formal dynamical systems, including computer programmes.  In this
    context classical thermodynamics appears as a particular theory
    suited to a subset of all dynamical systems.  A statistical
    mechanics similar to the one derived with the microcanonical
    ensemble emerges from dynamical systems provided it contains, 1) a
    finite non-integrable part of its phase space which is, 2) ergodic
    at a satisfactory degree after a finite time.  The integrable part
    of phase space provides the constraints that shape the particular
    system macroscopical properties, and the chaotic part provides
    well behaved statistical properties over a relevant finite time.
    More generic semi-ergodic systems lead to intermittent behaviour,
    thus may be unsuited for a statistical description of steady
    states.  \\
    Following these lines of thought, in a second part non-extensive
    hierarchical systems with statistical scale-invariance and power
    law interactions are explored.  Only the virial constraint,
    consistent with their microdynamics, is included.  No assumptions
    of classical thermodynamics are used, in particular extensivity
    and local homogeneity.  In the limit of a large hierarchical range
    new constraints emerge in some conditions that depend on the
    interaction law range.  In particular for the gravitational case,
    a velocity-site scaling relation is derived which is consistant
    with the ones empirically observed in the fractal interstellar
    medium.
\end{abstract}


\end{frontmatter}

\section{The scope of classical statistical physics}

\subsection{The non-universality of  classical statistical mechanics}
The thermodynamical principles were sometimes believed to be
applicable to all ``sufficiently large'' physical systems,
including the whole Universe.  The principles of thermodynamics have
been so well verified in terrestrial conditions that the extrapolation
of universal validity was adopted by most scientists.

In fact this optimism had to be seen as exaggerated after, among
others, the pioneer works in astronomy of H\'enon (1961), Antonov
(1962), and Lynden-Bell \& Wood (1968) about the thermodynamics of
gravitating systems.  These works showed that well established rules,
such that specific heat is always positive, are not
verified by all large systems.  The awareness that the scope of
classical thermodynamics is restricted to a subset of physical systems
has spread slowly but steadily among physicists and chemists
(Lynden-Bell 1999).  In particular a large part of classical
thermodynamics is restricted to extensive systems.  Thus the old
assumption, made among others by Kelvin, that thermodynamics in the
present form applies to the whole Universe remains to be verified.  In
practise astronomers know well that classical statistical mechanics
does not work for gravitational systems, since they need to perform
intensive computer simulations of gravitating particles to predict
their average collective behaviour and evolution.  In contrast,
chemists and physicists can rely on thermodynamics to predict a large
range of possible collective states adopted by molecules.

Therefore it seems useful today to better characterise the effective
scope of classical statistical physics, and eventually to try to
extend it to more systems, including non-extensive systems and non
physical systems.  Such extensions toward non-physical systems have
been attempted many times, for instance in information theory (Shannon
1948), or for image processing.  But the assumptions made are not
better founded than the ones of statistical physics, therefore much
remains to be clarified.

On the other hand statistical tools are used in many different areas
of human activities to describe complex systems, natural inorganic
systems as well as organic or sociological systems.  Besides general
mathematical tools there does not seem to exist uniform statistical
principles similar to the ones of thermodynamics that can be applied
for all the possible systems.

We will suggest in the following that the main characteristics that
systems need for applying statistical tools similar to thermodynamics
is a chaotic microscopic dynamics leading to a fast ergodic behaviour,
which may however still be subject to global constraints associated
with conservation laws.  Sufficiently chaotic dynamical systems that
nevertheless are constraint by symmetry invariants explore a subset of
their phase space in a sufficiently good ergodic fashion.  This allows
both to recognise general symmetries coming from the phase space
restriction, and to use statistics appropriately over the almost ergodic
domain.

\subsection{Thermodynamics is not necessarily fundamental}
The principles of thermodynamics were elaborated during the 19$^{\rm
  th}$ century and found empirically to be useful for particular
systems, such as thermal machines, or diluted gases.  Later during the
20$^{\rm th}$ century the thermodynamical principles could be shown to
be consistent with a microscopic description based on classical or
quantum mechanics, provided some assumptions (ergodic principle,
extensivity, etc.) were introduced.  In textbooks these assumptions
are often not all explicitly stated, in particular that
thermodynamical variables must be either intensive or extensive. This
later rule has been sometimes viewed as an additional principle of
thermodynamics.

But some systems, such as turbulent flows, semi-chaotic mechanical
systems, or systems in a phase transition do not follow globally
thermodynamics; yet they display reproducible behaviours in a
statistical sense.  Such systems do not belong to systems describable
with the tools of statistical mechanics due to assumptions being
violated.  For example, some systems are out of mechanical or chemical
equilibrium, some others are non-extensive, some others are
insufficiently chaotic.

So the scope of thermodynamics must be viewed as restricted to
particular physical systems.  The key point to observe is that in
certain conditions it is possible to simulate thermodynamics on
computers, that is, on formal systems. This shows that thermodynamics
may emerge from purely abstract dynamical systems, so is not more
fundamental than the underlying dynamical system.  Molecular dynamics
using classical dynamics is today an important formal tool to
understand natural processes involving millions of atoms.  The pioneer
work by Fermi, Pasta and Ulam (1955) failing to reproduce on computer
the expected approach to thermal equilibrium by a chain of coupled
non-linear oscillators did precisely show that the chosen mathematical
model may also be in conflict with classical thermodynamics.  So one
can sometimes simulate thermodynamics with purely numerical models
based on ordinary differential equations, which are based for example
on classical mechanics; or, it may happen that particular classical
mechanical systems violate the usual thermodynamical laws.
Thus it seems clear today that classical statistical physics can
hardly be supposed to be more fundamental than the underlying
microdynamics of the considered systems.  For example the statistical
behaviour of a stellar cluster has nothing to do with the physical
nature of its stars, since the same behaviour can be simulated with
the classical dynamics of point particles on a computer.  Instead of
a fundamental theory, classical thermodynamics appears today as a
meta-theory applicable to a subset of all dynamical systems subject to
particular constraints.  Since thermodynamics can be abstracted from 
the underlying physics, thermodynamics emerges from the average
dynamics of certain chaotic systems with many degrees of freedom.

As consequence, when applicable thermodynamical quantities and their
properties, in particular entropy, should be derivable from the
underlying dynamics.  Contrary to quantities like time, space, and
mass, thermodynamical quantities do not need to be assumed to be
fundamental, since they can be represented on computer by properly
averaging the system dynamical quantities.

\subsection{Dynamical systems}
In the following we restrict our considerations on classical systems
for which the relationship between chaos and integrability on the
statistical behaviour is much better understood than in quantum
systems.  Also classical systems can be simulated on computers much
more easily than quantum systems, which helps forging a mental model
of their statistical behaviour.  Much progress has been achieved in
the last decades about understanding the correspondence of classical
chaos, thermodynamics and quantum mechanics (e.g. Gemmer et al. 2004),
but to not complicate the discussion we will restrict it below on
classical systems.

\begin{figure}[t]
\centering
\includegraphics[width=14cm]{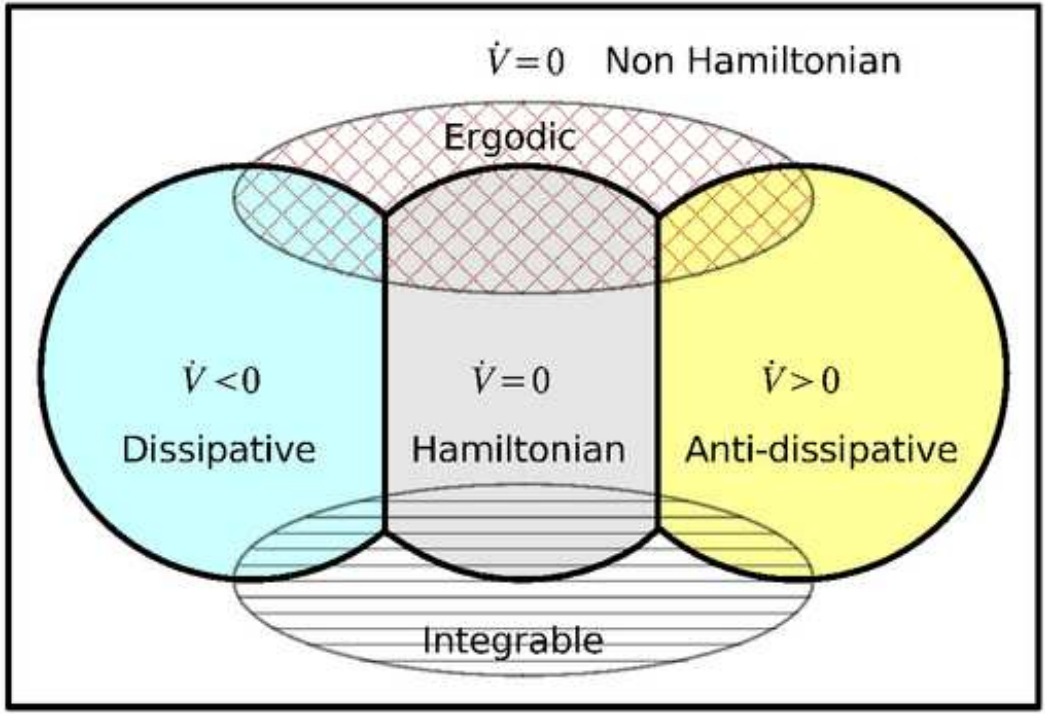}
\caption{The set of all dynamical systems includes the dissipative
  ($\dot V<0$) and anti-dissipative ($\dot V>0$) system subsets.
  Their complement is the subset of volume conserving ($\dot V=0$)
  systems, in which the Hamiltonian systems are a subset.  Integrable
  and ergodic systems are distinct subsets overlapping the previous
  subsets.  Classical statistical physics is restricted to the ergodic
  subset, while deterministic physics is restricted to the integrable
  subset.}
\label{fig:1}
\end{figure}

A general dynamical system is often defined by the phase space $\{\vec
z\}$, a set of real vectors in a $N$-dimensional Euclidean space,
generated by a set of first order ordinary differential equations
\begin{equation}
\dot{\vec z} = \vec F(\vec z, t) \ .
\end{equation}

The set of all dynamical systems is a too broad set to build
thermodynamics upon. Dissipative systems have attractors, therefore
their asymptotic behaviour is restricted to these attractors, which
may be as simple as a single point in phases space.  Richer attractors
are limit cycles, and still richer attractors, the strange
attractors, may have fractal dimension.  Reversing time in a
dissipative system produces an anti-dissipative system which tends
asymptotically to explore the boundaries of the allowed phase space.

Naturally some systems may have some dissipative phase space
coordinates and some other anti-dissipative coordinates.  Among
such systems the Hamiltonian systems form a special subset where the
phase space variables may be chosen in conjugate pairs in such a way
that each of them conserves its phase space area.  The symplectic
structure of phase space follows from the Hamiltonian $H(\vec p, \vec
q)$ and the particular form of Hamilton's equation of motion.
Hamiltonian systems still form a too broad set for classical
thermodynamics, since some of these systems have long range
interactions for which energy is not scaling proportionally to volume
or mass.  Others Hamiltonian systems are fully integrable, that is
decomposable into independent one-degree of freedom Hamiltonian systems
(consisting each of one pair of conjugate variables). Due to the area
conservation constraint, such systems cannot be chaotic.

However, for many Hamiltonian systems classical statistical mechanics
can already be used without the assumption of extensivity, as
discussed by Gross in this volume and elsewhere (e.g. Gross 2001), or
Padmanabhan for gravitating systems (1990).  This is the
microcanonical description, which however still requires the ergodic
hypothesis, which in turn requires a finite phase space volume.
Further, physical systems are modelled over a finite time over which a
statistical description makes sense. This often means that the
strength of chaos, measured by the rate of exponential divergence of
neighbouring trajectories, must be large enough to effectively
provides an sufficiently good ergodic behaviour over the physically
relevant finite time.

The Fermi-Pasta-Ulam (1955) oscillator model first showed that not any
mechanical system is ergodic, and subsequent works on dynamical
systems such as the Kolmogorov-Arnold-Moser (KAM) theorem provided
many more indications that most dynamical systems are partly chaotic,
or semi-ergodic.  Thus typical Hamiltonian systems are in general not
ergodic, therefore the ergodic hypothesis is not necessarily valid in
integrable or semi-ergodic systems. 

In integrable systems almost all the orbits are quasi-periodic,
therefore the initial correlations given by the initial conditions are
never erased.  Although exactly integrable systems are exceptional,
fully ergodic systems are also rare.  Generic Hamiltonian systems
possess phase space regions with local quasi-integrals, which are
nevertheless riddled by a web of resonances and chaos.  In such
semi-ergodic systems the accessible phase space over finite but long
``quasi-asymptotic'' times can be confined to a subset of the
non-integrable phase space that depends finely on the particular
initial conditions.  This leads to intermittent behaviours.  In such
systems the usual statistical mechanical tools are not appropriate to
describe steady states, because sudden large fluctuations are possible
after arbitrary long times.

In a mechanical description of a physical system the basic properties
that imprint global symmetries are the invariant functions of the
coordinates, the global integrals of motion.  Since in all time
independent mechanical systems the total energy function $E=H(\vec
p(t), \vec q(t))$ is the associated invariant, it is not astonishing
that energy plays a key role in thermodynamics.  Any invariant of the
system introduces a geometrical constraint in phase space that is
obeyed by the system, despite the presence of many more remaining
fully unconstrained degrees of freedom.

Here enters the crucial property that most systems where
thermodynamics works sufficiently well must have: most of the degrees
of freedom must be strongly chaotic, which means that beyond a
relaxation time the unavoidable and not accountable perturbations of
the rest of the Universe on the modelled system can no longer be
neglected. Beyond this time a deterministic description of the system
is impossible.  If the perturbations are uncorrelated with the
system degrees of freedom, then rapidly a strongly chaotic system will
have an equal probability to visit almost all points of the
non-integrable subspace of phase space.  Therefore the ergodic
principle in statistical mechanics only applies over the variables
belonging to the sufficiently chaotic degrees of freedom.  Then over
these variables averaging works well.  In contrast, each strict
integral of motion lowers the dimension of the available phase space
by one.

As an example of system with non-thermodynamical behaviour, for most
initial conditions a gravitational 10-body system dissolves in a few
crossing times, while for particular initial conditions like the
present solar system ones the same 10-body system may stay bound over
billions of crossing times.  The solar system is therefore a prime
case of semi-ergodic system on which classical thermodynamics does not
apply.  Over time-scale of a few million years only a small part of
the degrees of freedom behave in a chaotic way \citep{Laskar:1994}, indeed
the other degrees of freedom are constraint not only by the 10 global 
classical integrals of motions, but also by additional local 
constraints that act almost like additional integrals of motion
over very long finite time-scales.

A natural global quantity of a bounded mechanical Hamiltonian system
following the orbit $\left\{ \vec P(t), \vec Q(t)\right\}$ is the total
visited phase space volume
\begin{equation} 
  \Gamma =\lim_{t\to \infty}{1\over 2t}\int_{-t}^{t}\! dt'
  \int\!  d^N\vec q\, d^N\vec p ~
  \left[ \delta\!\left(\vec q -\vec Q(t') \right)
         \delta\!\left(\vec p - \vec P(t') \right)
  \right] \ .
\end{equation}
If the orbit were ergodic, almost all the points at constant value of
$E$ of $H(\vec p,\vec q)$ would be included in $\Gamma$.  This
constant $E$ volume can also be expressed as
\begin{equation}
\Gamma_E = \int \! d^N\vec q\, d^N\vec p ~
   \delta\! \left( E-H(\vec p,\vec q) \right)   \ .
\end{equation}
Every additional independent invariant function $I_k(\vec p,\vec q)$
restricts the available space space by one dimension in general
systems, by two dimensions in Hamiltonian systems due to the
symplectic constraint.  If the number of chaotic degrees of freedom
exceeds by far the number of invariant functions, the familiar
situation for gases, the total accessible phase space volume remains a
good approximation of the total phase space volume.  This quantity,
taken in the log, is, up to a constant, nothing else as Boltzmann's
original entropy $S = k_B \ln \Gamma_H$.  For a statistical mechanics
based on a well defined dynamics, such an entropy has a well defined
sense.  Without such a definition based on the particular system
microdynamics, entropy remains a non-measurable quantity with obscure
physical meaning.
\footnote{John von Neumann once advised Claude Shannon to call
  ``entropy'' the measure of information: ``No one knows what entropy
  really is, so in a debate you will always have the advantage''
  \citep{Tribus:1971}.}

Explicitely, if the Hamiltonian system has $N$ degrees of freedom,
$M\leq N$ global independent integrals $I_k(p_i,q_i)$ with values
$J_k$, and if the remaining accessible phase space is ergodic, then
the log of the accessible volume,
\begin{equation}
  \Gamma_{\rm ergodic} = 
\int \prod_{k=1}^M̀ \delta \left( J_k-I_k(\vec p, \vec q) \right) 
\, dq_i^N dp_i^N  \ ,
\end{equation}
can be used as the system microcanonical entropy, on which a
thermodynamics can be built.

\subsection{Generalisations}
From there it is straightforward to generalise these considerations
for a wider class of dynamical systems.  First, Hamiltonian systems
form a subset of volume preserving systems.  By taking a Cartesian
product of simpler volume preserving systems and combining the
variables by regular functions one can build more general coupled
dynamical systems with an odd or even number of variables that are
still volume preserving, and possess a number of global integrals of
motion.  Like for Hamiltonian systems the remaining accessible phase
space however may or may not be effectively ergodic due to chaos.  In
the first case the entropy can be defined and be used to built upon a
microcanonical thermodynamics.

For dissipative dynamical systems $\dot{\vec z}=\vec{F}(\vec{z})$ with
orbit $\vec Z(t)$, the asymptotic attractors may still represent a
non-trivial invariant subset of phase space which is eventually
ergodic.  For such systems the asymptotic thermodynamics is still a
well defined procedure, just taken in the asymptotic limit of large
time,
\begin{equation}
  \Gamma_{\rm ergodic} = 
\lim_{t_2\to \infty} \lim_{t_1 \to \infty} 
{1\over t_2} \int_{t_1}^{t_1+t_2}\!  dt' 
\int \! d\vec z ~
\delta\! \left( \vec z -\vec Z_i(t') \right)  \ .
\end{equation}
Actually, most steady natural systems can be supposed to follow this
route.  Starting from non-equilibrium states they converge toward 
asymptotic average steady states due to dissipation, and are maintained
on their respective final attractors by global integrals.

\subsection{The reason for extensive variables}
An important point to explicit is the reason why thermodynamical
variables in the canonical description are supposed to be either
intensive, i.e., position invariant, or extensive, i.e., scaling
linearly with the spatial volume $V$.  For some reason, rarely
discussed in depth, the thermodynamical variable $X_i$ of a system
with many degrees of freedom scales as
\begin{equation}
X_i \propto V^\sigma\ ,
\end{equation}
where $\sigma=0$ or $\sigma=1$.  To assume such scaling relations is a
crucial step leading to the usual characteristics of classical
thermodynamics.  This symmetry allows to describe extensive systems
with uniform prescriptions encouraging the belief that extensive
thermodynamics is universally applicable.

Let us apply the considerations of previous section to explain why
systems with a large number of degrees of freedom tend toward uniform
density above some microscopic scale.  The smooth average distribution
of particles is necessary to describe macroscopic variables as
differentiable, which is then required to allow the use of
differential equations for describing globally out of equilibrium
fluids only locally in thermal equilibrium.  In such
systems the particle Lagrangian $\mathcal{L}$ for the space
coordinates $\{ \vec x_i \}$ takes the usual form where $T$ is the kinetic energy and $U$ the potential energy,
\begin{equation}
\mathcal{L} = T(\dot{\vec x}_i) - U({\vec x}_i), \quad i=1,\ldots N. 
\end{equation}
In systems where the particle interaction energy is negligible with
respect to their kinetic energy, $|T|\gg |U|$.  The Lagrangian becomes
then independent on the space coordinate $\vec x_i$.  Therefore any
integral of the system and subs-system should adopt the same space and
orientation invariance: the system of particles must tend toward a
uniform spatial distribution of the particles.  In velocity space the
particles are constraint by a finite total energy, a spherical
symmetry of the distribution function in velocity space with a finite
second moment follows; for distinguishable particles the Maxwell-Boltzmann
distribution follows.  Therefore the uniform smooth mass density in
classical gases and fluids in statistical equilibrium follows from the
corresponding \textit{invariance of the Lagrangian} and from the
chaotic nature of motion of particles.
\footnote{Although the perfect gas model neglects formally the
  interaction term in the Lagrangian, collisions are essential in real
  gas for making the molecular motion chaotic.  Otherwise the chaos
  would depend entirely on the boundary conditions.}
The extensivity property follows then from the microscopic dynamical
properties of chaotic particles subject to a mutual interaction energy
small in regards of their kinetic energy.  The intensive
variables, for example the mass density $\rho$, derive from the ratio
of two extensive variables, such as the ratio of the mass and the
volume.

At this point it is clear that when the interaction potential energy
is not negligible, at sufficiently low temperature, the rules of
classical thermodynamics can break down.  Indeed, as stressed for
example by \cite{Lynden-Bell:1999}, at sufficiently low temperature a
phase transition occurs, the microcanonical specific heat becomes
negative.  The system is then thermally unstable, the homogeneity
invariance breaks down due to coexisting phases.

At still lower temperature the interaction energy in the Lagrangian
dominates; a crystal may form, in which case most of the degrees of
freedom become frozen.  The continuous translational invariance is
then broken and replaced by a discrete lattice invariance.  The
remaining small motions of an ensemble of molecules around the lattice
energy minima form again an extensive dynamical system which can be
described with usual statistical physics at the condition that the
small motions around the lattice minima are nevertheless chaotic, 
unlike the Fermi-Pasta-Ulam oscillators.

\section{Non-extensive systems}
However many cold systems with strong interaction potential do not
develop regular lattices, among them glasses, or the gravitational
$N$-body systems.  Since the statistical behaviour of gravitational
systems is observed to display similar properties across galaxies and
the Universe, a statistical description of them can nevertheless be
expected.

An important additional symmetry in systems of particles such as
gravitational systems is that the long range interaction term is scale
invariant. In other terms the potential energy follows the symmetry of
$p$-homogeneous functions,
\begin{equation}
U (\lambda \vec x_i) = \lambda^p U(\vec x_i) \, , 
\end{equation}
where $\lambda$ is any real positive number, and $p$ is the power law
exponent of the 2-particle potential; $p=-1$ for gravitational
systems.  For $p$ sufficiently negative the potential is short ranged,
so we should expect to recover perfect gas behaviour at sufficiently
negative $p$.

So if we follow the above reasoning, a system of strongly interacting
particles might display statistical behaviour following different
symmetries than the ones adopted in extensive systems: instead of
spatial invariance, systems may be characterised by scale-invariant
properties.  A priori nothing obliges us to assume a smooth,
differentiable average particle distributions.  Distributions in the
mathematical sense, such as fractal or statistically scale-invariant
distributions are actually suggested by many physical systems, as
popularised by Mandelbrot (1982).  Such distributions are in general
non-differentiable, but can still be represented by an ensemble of
points, i.e., a superposition of Dirac distributions.

So the problem that we want to address is to describe statistically such
systems without adopting the usual assumptions of thermodynamics, but
using as much as possible the known constraints from the microdynamics.
Clearly we do not want to start with the assumption of extensivity,
but rather take the symmetries of the Lagrangian as a guide to
describe it in a statistical way.

Instead of requiring microscopic extensivity which justifies smooth
microscopic distributions just above the mean-free path scales, we
want to allow instead scale-invariant hierarchical systems, since such
systems appear frequently in nature.  The consequence is that the
usual mathematical tools mostly based on differential equations are no
longer suited.

The main general statistical tool of particle dynamics is the virial
theorem, which applies to any, smooth or non-smooth, distribution of
particles. In the literature the virial theorem is often derived by
averaging the fluid equations of motions.  This approach assumes from
the start that the distribution function is regular.  However the more
general route not requiring regularity is to use the Lagrange-Jacobi
identity valid for any set of point mass particles.

The polar semi-moment of inertia $I$ about the origin of a system of
particles of mass $m_i$
\begin{equation}
I = {1 \over 2} \sum_i m_i \vec x_i^2 \, ,
\end{equation}
characterises the system average mass extension in space.
Differentiating $I$ with respect to time, the acceleration of this global
scalar quantity is seen to be linked to the system kinetic energy and
the total moment of radial forces, where ${\vec F}_i = m_i \ddot {\vec
  x}_i$,
\begin{equation}
\ddot I = 
\sum_i m_i \dot {\vec x}_i^2 + 
\sum_i {\vec F}_i \cdot {\vec x}_i \,  .
\end{equation}
If the forces derive from a sum of $p_k$-homogeneous potentials
$U_k$, 
\begin{equation}
\ddot {\vec x} = - \sum_k \nabla U_k({\vec x}_i)  \, , 
\qquad 
\textrm{where}
\qquad 
U_k (\lambda \vec x) = \lambda^{p_k} U_k(\vec x) \, , 
\end{equation}
then the total moment of force is proportional to a sum of potential
energies weighted with the homogeneity indexes $p_k$, thus
\begin{equation}
\ddot I = 
2  T({\vec x}_i^2) - \sum_k p_k U_k ({\vec x}_i) \,  .
\end{equation}
This is the Lagrange-Jacobi identity, slightly generalised with a
sum of homogeneous potentials.

If the system is confined by a pressure at the boundary, this pressure
has the same effect that a uniform negative kinetic energy distributed
inside the system volume $V$.  Since a non-relativistic kinetic
pressure amounts to $2/3$ of the kinetic energy density, we can write
in virial equilibrium ($\ddot I = 0$) that
\begin{equation}
2  T({\vec x}_i^2) - \sum_k p_k U_k ({\vec x}_i)  =
3P_{\rm ext} V  \,  .
\end{equation}
%

\section{Hierarchical systems}

\begin{figure}[t]
\centering
\space{1cm}
\includegraphics[height=8cm]{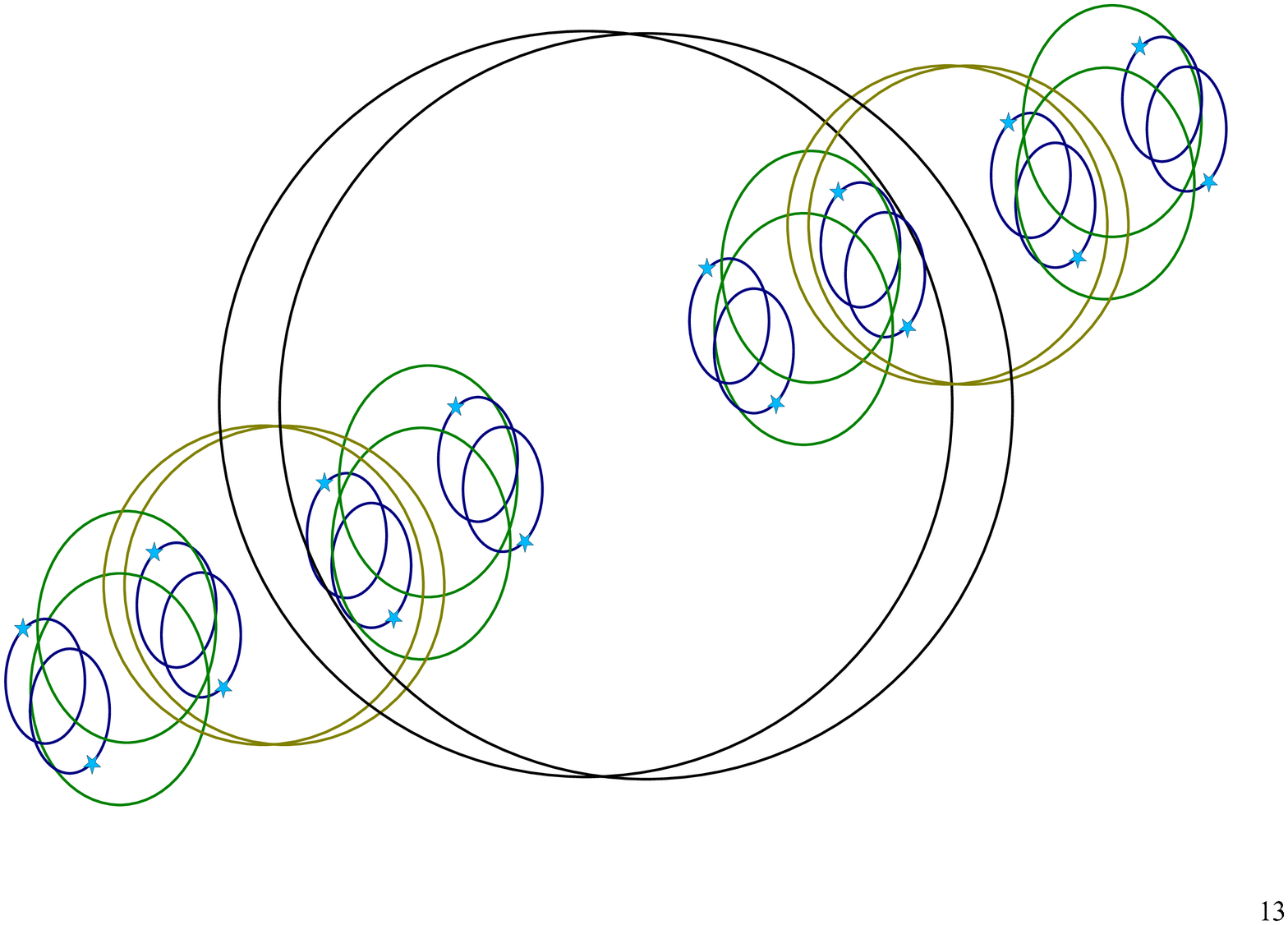}
\caption{A hierarchical $2^4$-body system.}
\label{fig:2bodies}
\end{figure}

\subsection{The hierarchical $N^L$-body problem}

The gravitational $N$-body problem has played a central role in the
development of classical mechanics and the theory of dynamical
systems.  It was seen for over 2 centuries as a prime example of the
deterministic physics developed by Newton and followers.  The main
reason is, first, that the 2-body case is fully integrable, so
computable over an arbitrary long time, and, second, that the planets
move on orbits that are perturbations of the 2-body problem.
Therefore the solar system remains well computable over centuries. But
the works on the 3-body problem by Brun and Poincar\'e in the late
19$^{\rm th}$ century showed that the general $N$-body problem was of
a different kind, not so easily computable.  The notion of exponential
divergence of nearby trajectories leading to the modern notion of
chaos came then into existence.

Hierarchical $N$-body problems have been rarely studied (e.g., Nugeyre
\& Bouvier 1981).  Yet such systems may provide others configurations
than the planetary systems that are close to be integrable.  Let us
consider a bound 2-body gravitational system.  If now we replace each
body by an identical but scaled down 2-body problem at its centre of
mass, and then recursively over a number $L$ of levels, then we have
built a hierarchical $2^L$-body problem (Fig.\ref{fig:2bodies}).
This system is close to be integrable, since each 2-body subsystem is
perturbed by the inner and the outer other 2-body systems.  In order
to avoid possible resonances we have the freedom to choose the scale
ratio is such a way as to avoid rational ratios of orbital periods.
Then in the limit of a large scale ratio the respective perturbations
by the other bodies on each two body system is small, eventually
negligible for at least the finite time we wish to consider such a
system.

\subsection{Geometrical relations}
Let us first enumerate some simple but little known geometrical
properties of mass particles arranged in a hierarchical way (Pfenniger
\& Combes 1994).

As first hypothesis we consider a system of mass $m_L$ made of $N$
clumps of mass $m_{L-1}$, made themselves of $N$ sub-clumps $m_{L-2}$,
and so on over a number of levels.  At any level $L>0$,
\begin{equation}
m_L = N m_{L-1}  \ .
\label{equ:M}
\end{equation}
In a regular hierarchical system the ratio of masses between adjacent
levels is supposed to scale with spatial size according to a power law
$d$,
\begin{equation}
{m_L \over m_{L-1}} = \left( {r_L \over r_{L-1}}\right)^d  \ .
\label{equ:MR}
\end{equation}
Since mass is positive, its cumulative value within a sphere centred
on a mass point must monotonously increase with the sphere radius, so
$d \ge 0$.  In a 3-dimensional space, the positive cumulative mass
cannot grow faster than volume, therefore $d \le 3$.

As consequence there is a direct relationship between the adjacent
level size ratio, $N$ and $d$,
\begin{equation}
{r_L \over r_{L-1}} = N^{1/d}  \ .
\label{equ:RND}
\end{equation}

It follows straightforwardly that the average mass density scales
as
\begin{equation}
{\rho_L \over \rho_{L-1}} =  
   \left( {m_L/ r_L^3 \over m_{L-1} / r_{L-1}^3}\right) = 
   \left( {r_L \over r_{L-1}}\right)^{d-3}  \ ,
\end{equation}
which means that the average density increases at small scale
for all permitted $d < 3$.
  
In contrast, the projected mass density $\Sigma_L$ scales as
\begin{equation}
{\Sigma_L \over \Sigma_{L-1}} =  
\left( {m_L/ r_L^2 \over m_{L-1} / r_{L-1}^2}\right) = 
\left( {r_L \over r_{L-1}}\right)^{d-2}  \ .
\end{equation}
Thus the critical exponent $d=2$ separates the projection of fractal
objects of different $d$ (see also Falconer 1990).  The immediate
consequence for astronomical observations restricted by projected
views of cosmic structures is that if a fractal object has $3 \ge d >
2$ then the sky is well filled by it, like common fog does.  However,
when $d < 2$ most of the mass is projected in small spots, leaving
most of the sky empty.  Therefore a fractal distribution of galaxies
with $d<2$ (Coleman \& Pietronero 1992) is a possible (partial)
explanation of the blackness of the Universe, the famous de
Ch\'eseaux-Olbers paradox.

\subsection{Dynamical relations}

Hierarchical distributions of gas are observed in the interstellar
medium without a clear understanding of the overall physics.  It seems
clear that in a galaxy gravity plays a non-negligible role at most
scales.  Typically the interstellar gas fills the volume of a galaxy
in a very inhomogeneous way well described by fractal geometry.  The
matter is fragmented in clumps themselves composed of sub-clumps over
several order of magnitudes in scale.  At each scale of a galaxy mass
condensations are rather close to virial equilibrium from sizes
comparable to the whole galaxy down to stars, passing by various
interstellar clouds of intermediate sizes.  Between the largest galaxy
size and the smallest stellar size, a range of order $10^{12}$ is
covered.
  
Generalising the gravitating case to other power law potentials, we 
adopt the potential form between mass particles,
\begin{equation}
\Phi(r) =  {G M \over p}\, r^p \ , 
\label{equ:Phi}
\end{equation}
with $p\neq 0$.  The limit $p\to 0$ is well behaved and provides a
logarithmic potential, that we do not discuss separately.  For
attracting interactions the coupling constant $G$ is positive.
The gravitational case occurs for $p=-1$.

\begin{figure}[t]
\centering
\vspace{2cm}
\includegraphics[height=6cm]{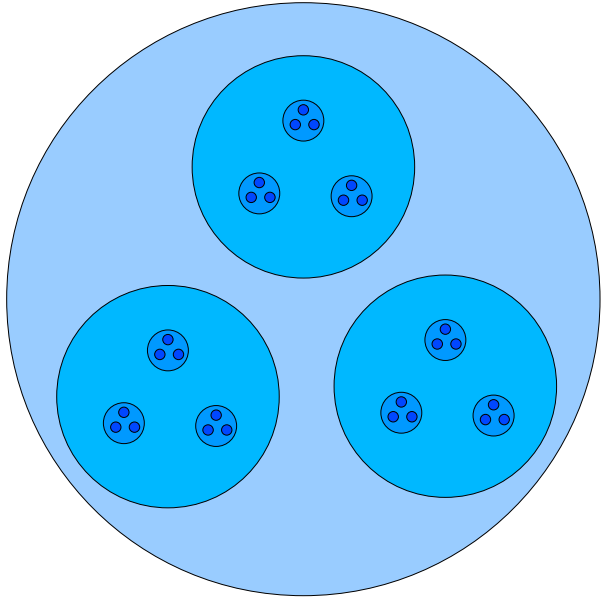}
\caption{The mass hierarchical model over 4 levels with $N=3$.}
\label{fig:2}
\end{figure}

Suggested by the system hierarchical organisation, we
\textit{approximate} the potential energy $U_L$ at level $L$ by
retaining the first term of a multipolar expansion,
\begin{equation}
U_L = N U_{L-1} +
\frac{G}{p} M_L^2 r_L^p \left( 1 + \alpha \frac{r_{L-1}}{r_L}\right)\  .
\label{equ:U}
\end{equation}
The potential energy at a given level consists of the summed
sub-clumps potential energies and the potential energy due to the
interaction between the sub-clumps.  The term containing $\alpha$
represents the first correction taking into account tidal
interactions.  Since the scale ratio is constant, the term in
parentheses is constant and can be absorbed in a new coupling constant
$G'$.  We will see that the main results do not depend on the {\it
  value\/} of $G$ (indeed scale invariance effects depend on the power
law exponent $p$).  For parameters suited to the interstellar medium
($D\approx 1-2$, $N=5-8$, Scalo 1990) the approximation (\ref{equ:U})
has been checked to be accurate at the percent level.
 
The third hypothesis is of statistical character.  Thermodynamics can
not be used since it excludes long-range interactions.  The only
remaining statistical tool is the virial theorem, more precisely the
Lagrange-Jacobi indentity.  If a scale invariant system finds a
statistical equilibrium at each level, then the semi-moment of inertia
acceleration $\ddot I$ of different sub-clumps vanishes in average,
such that at each level,
\begin{equation} 
2T_L - p U_L = 3 P_{L+1} V_L \, , 
\label{equ:VIR}
\end{equation}
where $T_L$ is the total kinetic energy summed up level $L$,
$P_{L+1}$ is the outer kinetic pressure, and $V_L \sim r_L^3$ the
clump volume.  Here the outer pressure is purely kinetic: it is given
by $2/3$ of the kinetic energy density outside the sub-clumps.
Approximately,
\begin{equation}
P_L = \frac{2}{3} {T_L - N T_{L-1} \over V_L - N V_{L-1}}, 
\label{equ:P}
\end{equation}
which takes into consideration that most of the time distinct clumps do
not overlap.

\subsection{Explicit solution}
The above recurrences for $M_L$, $V_L$, $U_L$ and $T_L$ in
Eqs.~(\ref{equ:M},\ref{equ:MR},\ref{equ:U},\ref{equ:VIR},\ref{equ:P})
can be solved with little algebra exactly in finite terms, as detailed
below.

First we define a few convenient terms, and state simple constraints.
The lowest level virial ratio,
\begin{equation}
\beta \equiv \frac{pU_0}{2T_0}
\end{equation}
is obviously constraint from Eq.~(\ref{equ:VIR}) divided by $2T_L$ in
the interval $0\leq \beta \leq1$, because $T_L$, $pU_L$, $V_L$, and
$P_L$ are all positive.  When $\beta\to 0$ the lowest level is
confined by outer pressure, when $\beta\to 1$, by internal attraction.

The lowest level energy is $E_0 = T_0 + U_0$.  The marginally bound 
state occurs when $E_0=0$.  Such a state is possible only when 
\begin{equation}
0 < -p/2 < 1 \, .
\end{equation}

The scale ratios 
\begin{equation}
x\equiv \frac{r_L}{r_{L-1}}, \quad {\rm and}\quad X \equiv x^L,
\end{equation}
are by definition $>1$, since as shown in Eq.~(\ref{equ:RND}), $x =
N^{1/d}>1$.

In terms of $x$ and $X$ we have,
\begin{equation}
\frac{m_L}{m_0} =  X^d, \quad {\rm and} \quad \frac{V_L}{V_0} =  X^3. 
\end{equation}
It just means that the mass is proportional to the power $d$ of the
size, while the volume to the power 3.  Incidentally, the factor
$X^{d-3}$ is the density ratio $\rho_L/\rho_0$.

Eq.~(\ref{equ:U}) is a linear recurrence relation for $U_L$ that has a
simple solution:
\begin{equation}
\frac{U_L/m_L}{U_0/m_0} = \frac{(Xx)^{d+p}-1}{x^{d+p}-1}
\, , \qquad L\geq 0\, .
\label{equ:Usol}
\end{equation}

Some more effort is required to solve Eq.~(\ref{equ:VIR}) by
substitutions and induction.  We obtain,
\begin{equation}
\frac{T_L /m_L}{T_0/m_0} = X^{3\!-\!d} +
 {\beta \over x^{d+p}\!-\! 1} 
 \left[ 
  { (Xx)^{3 -\!d}\! -\! (Xx)^{d+\!p} \over x^{3-d} - x^{d+p} }
 + { X^{3 -\!d}\! - \!X^{d+\!p} \over {1\over x^{3-d}} - {1\over x^{d+p}}} - 1
 \right] .
\label{equ:Tsol}
\end{equation}
The equivalent of the kinetic pressure at each level is proportional 
to the square of the velocity dispersion $v_L$ given by,
\begin{eqnarray}
\frac{1}{2} m_L v^2_L &\equiv& T_L \,  , \qquad \qquad \quad  L = 0 \, , \\
                      &\equiv& T_L - N T_{L-1}\,  , \quad L\geq 1 \, , 
\label{equ:v2}
\end{eqnarray}
which means that the kinetic pressure is the difference between two
adjacent levels of the summed kinetic energies down to the lowest
level.  Replacing with Eq.(\ref{equ:Tsol}), we obtain,
\begin{equation}
\frac{ v^2_L }{ v^2_0 } = 
\left(x^{3-d}-1\right) \bigg[ \left( \frac{X}{x}\right)^{3-d}
	-\beta \frac{ X^{3-d}-X^{d+p} } {x^{3-d} - x^{d+p}} \bigg]\,,
        \quad L\geq 1 \, .
\label{equ:v2sol}
\end{equation} 

\subsubsection{Physical domain of $p$ and $d$}
The above hierarchical model makes sense in the intervals $0<d\leq3$,
for geometrical reason, and $p<0$ if we wish to consider interactions
of attractive type not growing too fast with distance.  As already
understood by Newton, even the gravitating case $p=-1$ leads to
conceptual problems when $d=3$ (the force at any point in a uniform
infinite medium is a sum of cancelling diverging forces).

However, even in this restricted domain not any combination of $p$ and
$d$ is physically feasible.  One has to ensure that the positive
quantities are indeed positive, and do not diverge to excessive 
values. 

First one can easily check in Eq.[\ref{equ:Usol}] that $U_L/U_0$ does
not change sign because $X>x>1$, and at $d+p=0$ the behaviour is
regular.

The kinetic energy and the squared velocity (or the pressure) should
remain positive and finite.  In fact, by looking at Eq.[\ref{equ:v2}]
ones sees that it suffices to require the positivity of the squared
velocity at each level to ensure that the summed kinetic energy is
also positive.

\begin{figure}[t]
\centering
\vspace{0.cm}\hspace{0cm}
\includegraphics[height=14cm]{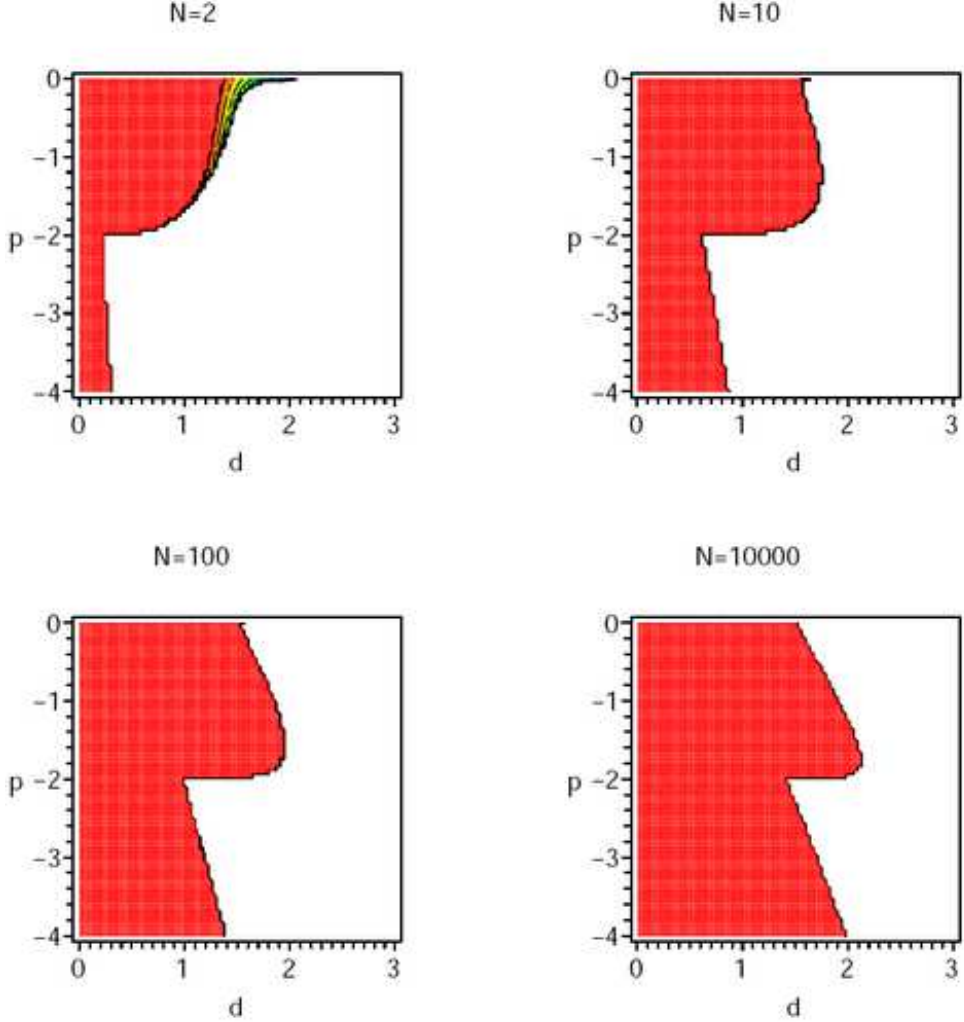}
\caption{The ratio $v_{L}^2/v_0^2$, for $l=12$, in the plane $(d,p)$
  takes either very large values exceeding $10^4$ in the shaded
  regions at low $d$, or negative values in the white region at large
  $d$, or physical values in the narrow intermediate region, for
  different $N$.  In all cases the low level boundary condition is
  $\beta = \min(1, -p/2)$, which produces the discontinuity at
  $p=-2$.}
\label{fig:v2v0}
\end{figure}

A denominator vanishes in Eq.[\ref{equ:Tsol}] when $d+p \to 0$.  But
more important, when $3-d \to d+p$ other denominators in
Eqs.[\ref{equ:Tsol}],[\ref{equ:v2sol}] vanish with corresponding
numerators containing cancelling powers of $X$.  This latter
singularity is crucial because it means that as the number of levels
of the system increases, a shrinking domain of parameter space is able
to correspond to accessible physical states.  As shown in
Fig.(\ref{fig:v2v0}), on one side of the singularity $v^2_L$ is
negative and on the other side it takes extremely large values, such
that the only physical sensible region is \textit{within} the
singularity as $L\to\infty$.

Since $v^2_L$ can change its sign, we obtain a limit of physically
admissible cases.  Solving Eq.[\ref{equ:v2sol}] for $\beta$ we obtain
a constraint for $\beta$, noting $\alpha = (d+p)-(3-d)$,
\begin{equation}
  v^2_L\geq 0 \quad \Rightarrow \quad \beta \leq 
  \frac{1-x^{\alpha}}{1-X^{\alpha}}.
\end{equation}
Since $X>x>1$, the critical value $\beta_{\rm crit}$ is in the
interval $(0-1)$ only when $\alpha>0$.  To not be subject to this
restriction on $\beta$, $d$ must obey
\begin{equation}
d \leq \min \left(3,{3-p \over 2} \right) \ . 
\end{equation}
For hierarchical gravitating systems with $d>2$, a range of $\beta<1$
is not allowed.  At the lowest level an outer pressure is required.

\subsection{Large $L$ limit}
In ordinary systems the thermodynamical limit means that the boundary
conditions, the ``box'', takes an infinite dimension, $V\to \infty$
and $M \to \infty$, but $M/V \to \rm const < \infty$, or the number of
particles $N$ tends toward infinity at constant average density.  The
mass and volume per particle are kept constant.  In other cases,
particularly relevant in gravitating systems such as stars in which
local thermodynamics still works, the total mass and volume is taken
as fixed, but the number of particles is increased to infinity by
decreasing the mass $m$ and volume $v$ per particle to zero, keeping
$m/v$ constant. 

In hierarchical systems, one can also increase the number of
particles, or degrees of freedom, toward infinity at either the large
or small scales, by increasing the number of levels $L$.  The boundary
conditions, fixing typically the temperature in a box, can be replaced
by the conditions at either small or large scale depending on the
problem.

Here we take the analogue of a ``thermodynamical limit'' by $L\to
\infty$, or $X = x^L\to \infty$.  Then the shrinking of the range of
physical solutions compatible with the virial theorem leads to the
constraint,
\begin{equation}
d =\frac{3-p}{2-\frac{\ln(1-\beta)}{\ln N}}.
\label{equ:dnew}
\end{equation}
Equivalently, one can use
\begin{equation}
\beta  = 1 - x^\alpha \, , \qquad \alpha = (d+p)-(3-d) \ .
\label{equ:betanew}
\end{equation}
For example for $p=-1$, $d<2$ in any case. For systems not too
confined by the outer pressure ($1/2<\beta<1$) and $N>5$, we have $d <
1.7$, while systems with more outer pressure (``pressure confined
clouds'') increase $d$ up to 2.  {\it Therefore, we conclude that
  hierarchical gravitating systems in statistical equilibrium can
  indeed exist, but with $d<2$.}

\subsection{Velocity scaling}
In the large $L$ limit, when using Eq.[\ref{equ:betanew}] the
scale-velocity dispersion in Eq.[\ref{equ:v2sol}] takes the exact
scale-free form
\begin{equation}
\frac{ v^2_L }{ v^2_0 } = 
\left(1-{ 1\over x^{3-d}}\right) X^{d+p} \, , \quad L\geq 1 \, .
\label{equ:v2solnew}
\end{equation} 
In other words
\begin{equation}
 v \propto r^{\kappa} \, , \qquad \textrm{where} \qquad 
\kappa\equiv {d+p \over 2}\, . 
\end{equation} 
For fractal ISM cold clouds with substantial ambient pressure, $\beta
\approx 1/2$, and with $N=5-10$ (Scalo 1990), we expect $d \approx
1.7$ and $\kappa \approx 0.35$, which is compatible with Larson's
(1981) size-linewidth relationship.

\section{Conclusions}

Once one realises that despite its enormous success, classical
statistical mechanics is not universal, but mostly applicable to
extensive strongly chaotic systems, a huge field of investigations
opens up.  Classes of systems, including formal or non-physical
systems, may eventually be found suited for a statistical description
of their steady states provided their dynamics is sufficiently chaotic
to allow the use of an ergodic hypothesis.  However, many semi-ergodic
systems exist with intermittent behaviour for which there is little
motivation to expect that a statistical description of steady states
is appropriate.

The non-extensive gravitational systems are the rule in astrophysics,
and many observational evidences exist that suggest that gravitational
systems may follow statistical rules.  For example laws such as the
$r^{1/4}$ radial density profile of elliptical galaxies, or the
exponential profiles of stellar galactic disks are well known, and
reproducible in computer simulations, but not fully explained by a
statistical model.  

Here as a illustrating example we have developed a statistical
description of $N^L$ hierarchical systems subject to power law
interaction without using concepts of classical statistical mechanics.
Using the virial theorem instead one can already find new constraints
that emerge for a large number of bodies.  As application, Larson's
empirical velocity scaling relationship in molecular clouds is
retrieved.

\subsection*{Acknowledgements}
I thank the organisers for this stimulating, timely and useful meeting
bridging different fields.


\end{document}